\documentclass[prd,floatfix,preprintnumbers]{revtex4}
\usepackage{amssymb}

\usepackage{amsmath}
\usepackage{graphicx,epsfig}
\usepackage{color}

\begin{document}

\title{Constraining Density Fluctuations
with Big Bang Nucleosynthesis
in the Era
of Precision Cosmology}
\author{John D. Barrow}
\affiliation{DAMTP, Centre for Mathematical Sciences, Cambridge University, Cambridge CB3
0WA, UK}
\author{Robert J. Scherrer}
\affiliation{Department of Physics and Astronomy, Vanderbilt University, Nashville, TN
~~37235}

\begin{abstract}
We reexamine big bang nucleosynthesis with large-scale baryon density
inhomogeneities when the length scale of the density fluctuations exceeds
the neutron diffusion length ($\sim 10^7-10^8$ cm at BBN), and the amplitude of the fluctuations is
sufficiently small to prevent gravitational collapse. In this limit, the
final light element abundances can be determined by simply mixing the
abundances from regions with different baryon/photon ratios without
interactions. We examine gaussian, lognormal, and gamma distributions for
the baryon/photon ratio, $\eta $. We find that the deuterium and lithium-7
abundances increase with the RMS fluctuation in $\eta $, while the effect on
helium-4 is much smaller. We show that these increases
in the deuterium and lithium-7 abundances are a consequence
of Jensen's inequality, and we derive analytic approximations
for these abundances in the limit of small RMS fluctuations.
Observational upper limits on the primordial
deuterium abundance constrain the RMS fluctuation in $\eta $ to be less than 
$17\%$ of the mean value of $\eta $.  This
provides us with a new limit on the graininess of the early universe.
\end{abstract}

\maketitle

\section{Introduction}

The successful theory of big bang nucleosynthesis (BBN) remains one of the
major pillars of modern cosmology. While BBN once treated the baryon/photon
ratio, $\eta $, as the main quantity to be determined by comparing BBN
predictions with astronomical observations, the independent measurement of $
\eta $ by deduction from cosmic microwave background (CMB) observations has
led to a minimal BBN theory with no free parameters. Using the CMB values
for $\eta $, the predicted BBN abundances of deuterium (D) and $^{4}$He are
in excellent agreement with the observations, while the predicted $^{7}$Li
abundance remains a factor of three larger than the observationally-inferred
abundance; for recent reviews of BBN, see refs. \cite{Cyburt,Mathews,Pitrou}.

Given the excellent agreement between the predicted and observed D and $^4$He
abundances, any modification that alters the BBN predictions will
probably be sharply constrained. Many such modifications have been proposed,
but here we focus on one of the earliest to be investigated: inhomogeneities
in the density. Such inhomogeneities are irrelevant if (as in the case of
adiabatic fluctuations or temperature fluctuations \cite{gisler}) they leave 
$\eta $ unchanged \cite{Pitrou}. Here, we will be interested in the
different case of isocurvature fluctuations, for which $\eta $ varies from
one region of the Universe to another. There is a long history of
investigations of these kind of models \cite
{Harrison,Wagoner,Zeldovich,Epstein,Barrow,Yang,Copi,Leonard} and their baryon inhomogeneity may
reflect environmental non-uniformities or symmetry breakings at phase
transitions in the very early universe, long before BBN occurred \cite{BT}.
These investigators showed that it was possible to reconcile the observed D
abundance with a closure density of baryons if the baryons were
inhomogeneously distributed. During the 1970s there was interest in baryon
symmetric cosmologies which had led to the consideration of proton and
neutron diffusion effects during BBN \cite{omnes,combes}, and the need to
avoid matter-antimatter annihilation effects by requiring that baryon
inhomogeneities were larger than the neutron diffusion length at BBN. The
subsequent realisation that baryon number would not be conserved in a Grand
Unified Theory (GUT) led to a rapid loss of interest in baryon symmetric
cosmologies, but there remained an interest in diffusion effects.
An important effect arises when the length scale of the
fluctuations is longer than the proton diffusion length but shorter than the
neutron diffusion length; in this case neutron diffusion from high-density
to low-density regions will strongly modify the neutron-proton ratio \cite
{Applegate,Alcock}. This class of models has been explored in great
detail (see, e.g., Refs. \cite{Lara,Nakamura} and references therein for more
recent discussions).  When neutron diffusion
is important and the fluctuation amplitude is large, it is possible to produce
heavier elements than those normally considered in standard homogeneous BBN \cite{Applegate2,Boyd,Nakamura},
a result which provides additional observational signatures and constraints on such models.
Furthermore, fluctuations of sufficiently large
amplitude can lead to gravitational collapse, further modifying the standard
BBN scenario \cite{Sale,Gnedin,Jedamzik} and creating primordial black holes.

In this paper, we ignore these latter possibilities (and also the
possibility of inhomogeneities in fundamental ``constants" like $G$ \cite
{CBS} or the neutron half-life) and concentrate on the simplest case:
low-amplitude, large-scale fluctuations. In this case, BBN proceeds in the
standard way in independent regions with different values of $\eta $, and
the elements in these regions mix after BBN to give the observed abundances
today. There are two reasons why this is an opportune time to revisit this
scenario. First, observational limits on the primordial element abundances
have improved significantly since the earlier investigations cited above. In
particular, high-redshift quasar absorption systems allow high-precision
measurements of the primordial value of $\mathrm{D/H}$ (the number ratio of
deuterium to hydrogen) \cite{Cooke,Zavarygin}. Second, $\eta $ is no longer
a free parameter to be determined from the results of BBN calculations.
Instead, it is fixed by measurements of CMB fluctuations and then becomes a
well determined input parameter for BBN. This is particularly salient for
the case of BBN with density fluctuations, as much of the motivation for
these models was the possibility that they might allow a wider range of
values for $\eta $ than in the standard model, and previous work
concentrated on determining the allowed range for $\eta $ in these models.
Here, instead, we will be able to use a combination of a fixed value for $
\eta $, along with improved estimates of the primordial element abundances,
to place limits on the magnitude and type of allowed density fluctuations in
the early universe.

In the next section, we will briefly review the standard model for BBN and
discuss the observational limits on the element abundances that we
incorporate in this paper. In Sec. III, we will examine a variety of
statistical distributions for $\eta $ and calculate the corresponding
predicted element abundances using
a version of the Kawano nucleosynthesis code 
\cite{Kawano} with updated reaction rates \cite{Cyburt2}.
We will then use the observational limits to bound the
fractional RMS fluctuation, $\sigma $, in each case. We discuss our results
in Sec. IV.

\section{Standard BBN}

Consider the standard model for BBN \cite{Cyburt,Mathews,Pitrou}. For $
T\gtrsim 1$ MeV, the weak interactions inter-convert protons and neutrons,
maintaining a thermal equilibrium ratio: 
\begin{eqnarray}
n~+~\nu _{e} &~\longleftrightarrow ~&p~+~e^{-},  \notag
\label{weak-interactions} \\
n~+~e^{+} &~\longleftrightarrow ~&p~+~\overline{\nu }_{e},  \notag \\
n &~\longleftrightarrow ~&p+e^{-}~+~\overline{\nu }_{e},
\end{eqnarray}
while a thermal abundance of deuterium is maintained by 
\begin{equation}
n~+~p~\longleftrightarrow ~\mathrm{D}~+~\gamma .
\end{equation}
After the weak reactions drop out of thermal equilibrium at $T\sim 0.8$ MeV,
free-neutron decay continues until $T\sim 0.1$ MeV, when rapid fusion into
heavier elements occurs. Almost all of the remaining neutrons end up bound
into $^{4}$He, with a small fraction remaining behind in the form of
deuterium, tritium, and $^3$He, and some production of $^{7}$Li and $^{7}$Be, with the latter
decaying into the former via electron capture at the beginning of the
recombination era \cite{Sunyaev}. The element abundances produced in BBN
depend on the baryon/photon ratio $\eta $, which can be independently
determined from the CMB. We adopt a value of $\eta =6.1\times 10^{-10}$,
consistent with recent results from Planck \cite{Ade}. This value of $\eta $
yields predicted abundances of D and $^{4}$He consistent with observations.

Recent observational estimates of D/H include those of the Particle Data
Group (PDG) \cite{PDG}: D/H $= (2.569 \pm 0.027) \times 10^{-5}$, Cooke 
\textit{et al.} \cite{Cooke}: D/H $= (2.527 \pm 0.030) \times 10^{-5}$, and
Zavarygin et al. \cite{Zavarygin}: D/H $= (2.545 \pm 0.025) \times 10^{-5}$.
Here we will adopt the PDG estimate: 
\begin{equation}  \label{D/H}
\mathrm{D/H} = (2.569 \pm 0.027) \times 10^{-5}.
\end{equation}

The primordial $^{4}$He abundance, designated $Y_{p}$, is not so well
established. Izotov et al. \cite{Izotov} give $Y_{p}=0.2551\pm 0.0022$,
while Aver et al. \cite{Aver} give $Y_{p}=0.2449\pm 0.0040$. The PDG limit
is \cite{PDG} $Y_{p}=0.245\pm 0.003$. However, as we will see, the primary
limit on the models we will examine here comes from deuterium, rather than $^{4}$He,
so the observational limit on $^{4}$He will have little effect on
our results.

As noted above, both the D and $^{4}$He abundances are consistent with the
predictions of standard BBN with the CMB value for $\eta $, but this is 
\textit{not} the case for the $^{7}$Li abundance. The primordial lithium
abundance is estimated to be \cite{PDG} 
\begin{equation}
^{7}\mathrm{Li/H}=(1.6\pm 0.3)\times 10^{-10}.
\end{equation}
However, standard BBN with $\eta \sim 6\times 10^{-10}$ predicts a
primordial value for $^{7}$Li/H that is roughly three times higher than this
observationally-inferred value. (For this value of $\eta $, most of the
primordial $^{7}$Li is produced in the form of $^{7}$Be, which decays into
$^{7}$Li much later, as noted above). This discrepancy between the predicted
and observationally-inferred primordial $^{7}$Li abundances is called the
\textquotedblleft lithium problem," and it remains unresolved at present
(for a further discussion, see Ref. \cite{Fields}). Clearly, we cannot use
the $^{7}$Li abundance to constrain inhomogeneous BBN, since it is already
inconsistent with standard BBN. However, it will be interesting to see
whether the inhomogeneous models examined here can help to solve the lithium
problem.

\section{Inhomogeneous BBN}

Our modeling of inhomogeneous BBN will closely parallel that of Refs. \cite
{Copi} and \cite{Leonard}. We will assume isocurvature fluctuations, with $
\eta $ varying across different regions of the universe. We will take the
length scales of these fluctuations to be larger than the neutron diffusion
length, so that diffusion of neutrons relative to protons is not a
significant effect. Note that the Planck CMB measurements strongly constrain
isocurvature modes \cite{Ade2}, but there are many orders of magnitude
between the comoving neutron diffusion length at BBN ($\sim 10^7 - 10^8$ cm
at BBN \cite{Rehm}) and the smallest length
scales probed by Planck, so our model has a nontrivial range of application.
We assume further that the inhomogeneous element abundances are smoothed out
by post-BBN diffusion of all of the nuclei, so that the observable universe
ends up with a single, uniform abundance of each element. Treatments of
post-BBN element diffusion in the standard (homogeneous) case can be found
in Refs. \cite{Loeb,Pospelov}.

Our model can be entirely characterized by the distribution of $\eta $,
given by the distribution function $f(\eta )$ which specifies the fraction
of the universe with $\eta $ in the interval $\left( \eta ,\eta +\delta \eta
\right) $ at the time of nucleosynthesis. We will consider a variety of
choices for $f(\eta )$. Since $f(\eta )$ is a probability
distribution for $\eta $, it must integrate to unity: 
\begin{equation}
\int_{0}^{\infty }f(\eta )~d\eta =1.
\end{equation}
We assume that the inhomogeneities present at BBN are erased by diffusion
before recombination, so that the mean value of $\eta $ today (and at
recombination) is given by 
\begin{equation}
\overline{\eta }=\int_{0}^{\infty }f(\eta )~\eta ~d\eta .
\end{equation}
The final average element abundances, i.e., the abundances inferred from
present-day measurements, are mass-weighted averages of the element
abundances produced in the different regions. These are most easily
expressed in terms of the mass fraction $X_{A}$ of a given nuclide $A$, for
which we have 
\begin{equation}
\overline{X}_{A}=\int_{0}^{\infty }X_{A}(\eta )~f(\eta )~\eta ~d\eta /
\overline{\eta }.
\end{equation}
The factor of $\eta $ in the integral comes from the fact that the
abundances must be weighted by the baryon density in each region before they
mix. The only complication is that the deuterium and $^{7}$Li abundances are
expressed as number ratios relative to hydrogen, D/H and $^{7}$Li/H,
rather than as mass fractions, where the relationship between $\mathrm{A/H}$
and $X_{A}$ is given by 
\begin{equation}
\mathrm{A/H}=\frac{X_{A}}{AX_{H}},
\end{equation}
and the hydrogen mass fraction, $X_{H}$ is, to a good approximation, given
by $X_{H}=1-Y_{p}$, where $Y_{p}$ denotes the primordial mass fraction of $
^{4}$He. Then the mean value of $\mathrm{A/H}$ averaged over different
values of $\eta $ is 
\begin{equation}
{\overline{\mathrm{A/H}}}=\frac{1}{A(1-\overline{Y}_{p})}\int_{0}^{\infty
}X_{A}(\eta )~f(\eta )~\eta ~d\eta /\overline{\eta }.  \label{A/H}
\end{equation}
where $\overline{Y}_{p}$ is the mean primordial value of the $^{4}$He mass
fraction in a given inhomogeneous model. In practice, $\overline{Y}_{p}$
never varies a great deal from the homogeneous value, $Y_p$, for the models
considered here, so the correction given by including the factor
of $1/(1-\overline{Y}_{p})$ instead of $1/(1-{Y}_{p})$ in Eq. (\ref{A/H}) is negligible.

To perform this calculation, all that is needed is a choice for $f(\eta )$
and the values of the element abundances as a function of $\eta $ in the
standard (homogeneous) case. A variety of choices for $f(\eta )$ have been
investigated: unimodal distributions have included the gamma distribution \cite
{Epstein,Yang}, the lognormal distribution \cite{Barrow}, and the gaussian
distribution \cite{Copi}. We will examine the same set of
distributions here.

Taking $\overline{\eta }$ to be given by $\overline{\eta }=6.1\times 10^{-10}
$, consistent with the Planck results \cite{Ade}, we define the comparison
ratio, $\phi ,$ to be given by 
\begin{equation}
\phi \equiv \eta /\overline{\eta }=\eta /6.1\times 10^{-10},
\end{equation}
so that our earlier expressions become 
\begin{eqnarray}
\label{phicond1}
\int_{0}^{\infty }f(\phi )~d\phi  &=&1,\\
\label{phicond2}
\int_{0}^{\infty }f(\phi )~\phi ~d\phi  &=&1,
\end{eqnarray}
along with 
\begin{eqnarray}
\label{Xphi}
\overline{X}_{A} &=&\int_{0}^{\infty }X_{A}(\phi )~f(\phi )~\phi ~d\phi ,\\
\label{Aphi}
{\overline{\mathrm{A/H}}} &=&\frac{1}{A(1-\overline{Y}_{p})}\int_{0}^{\infty
}X_{A}(\phi )~f(\phi )~\phi ~d\phi.
\end{eqnarray}
Since our goal is to use current observations to bound the magnitude of the
fluctuations, we will express all of our results in terms of the RMS
fluctuation in $\phi $, with mean value $1$, given by 
\begin{equation}
\sigma ^{2}=\int_{0}^{\infty }f(\phi )~(\phi -1)^{2}d\phi ,
\end{equation}
and $\sigma $ corresponds to the RMS fractional fluctuation in $\eta $ for a
given model (defined by a choice of inhomogeneity distribution $f$).

We now express the gaussian, lognormal, and gamma distributions in terms of $
\phi $, parametrizing each one by $\sigma $. First, consider the gaussian
distribution with unit mean, given by 
\begin{equation}
\label{gaussian}
f(\phi )=\frac{1}{\sqrt{2\pi }\sigma }e^{-(\phi -1)^{2}/2\sigma ^{2}}.
\end{equation}
Note that this distribution has the well-known problem that it implicitly
allows negative values for $\phi $ (and therefore for $\eta $) which are, of
course, unphysical. However, we will confine our attention to sufficiently
small values of $\sigma $ ($\sigma <0.25$) that the negative values of $\phi 
$ correspond to $4-\sigma $ fluctuations and therefore have no significant
effect on the final results. (In effect, we are truncating our gaussian at $
\phi =0$). The second distribution we consider is the lognormal with unit
mean, given by 
\begin{equation}
f(\phi )=\frac{1}{\sqrt{2\pi }s\phi }e^{-(\ln (\phi )-\mu )^{2}/2s^{2}},
\end{equation}
where 
\begin{equation}
s=\sqrt{\ln (1+\sigma ^{2})},
\end{equation}
and 
\begin{equation}
\mu =-\ln (1+\sigma ^{2})/2.
\end{equation}
Our final test distribution is the gamma distribution with unit mean, given
by 
\begin{equation}
f(\phi )=\frac{\alpha ^{\alpha }\phi ^{\alpha -1}e^{-\alpha \phi }}{\Gamma
(\alpha )}
\end{equation}
where $\Gamma (\alpha)$ is the gamma function, and $\alpha $ is related to $
\sigma $ by 
\begin{equation}
\alpha \equiv 1/\sigma ^{2}.
\end{equation}

Since our calculations of the mean element abundances involve an integration
over $X_{A}(\phi)$, we use a version of the Kawano nucleosynthesis code 
\cite{Kawano} with updated reaction rates \cite{Cyburt2}
and a neutron lifetime of $\tau = 880.2$ sec \cite{PDG}
to calculate the element abundances as a function of $\phi$.  Our predicted element abundances
at $\phi=1$ ($\eta = 6.1 \times 10^{-10}$) are D/H = $2.592 \times 10^{-5}$, $Y_p = 0.247$, and
$^7$Li/H $ = 4.544 \times 10^{-10}$.  These abundances are in good agreement with the corresponding
values in Ref. \cite{Cyburt}, but somewhat discrepant (for deuterium and lithium) from those in
Ref. \cite{Pitrou}.  (The differences in the abundances predicted in Ref. \cite{Cyburt} and those in
Ref. \cite{Pitrou} are most likely
due to differences in the reaction rates used in the corresponding computer codes).

Since we can only calculate the element abundances at
discrete values of $\phi$, we divide the range in $\phi$ into logarithmic bins and change the integrals in Eqs. (\ref{Xphi}) and (\ref{Aphi}) into the corresponding sums: 
\begin{eqnarray}
\label{sumX}
\overline{X}_{A} &=&\sum_{i}X_{A}(\phi _{i})~f(\phi _{i})~\phi _{i}~\Delta
\phi _{i},\\
\label{sumA}
{\overline{\mathrm{A/H}}} &=&\frac{1}{A(1-\overline{Y}_{p})}\sum_{i}X_{A}(\phi
_{i})~f(\phi _{i})~\phi _{i}~\Delta \phi _{i}.
\end{eqnarray}
We now proceed to calculate the inhomogeneous element abundances. Although
our ultimate goal will be a limit on the magnitude of the density
fluctuations, this limit will necessarily depend on current observational
limits. To derive a result less prone to obsolescence, we will first
determine the general effect on the element abundances by calculating the
ratio, $R$, of each element abundance in the inhomogeneous case to the
corresponding homogeneous abundance as a function of $\sigma $. We will
denote this ratio by $R_{\mathrm{D}}$ for deuterium and $R_{\mathrm{Li}}$
for $^{7}$Li. For each test distribution, we evaluate Eq.
(\ref{sumA}) using the binned abundances calculated numerically as a function
of $\phi $ to derive $R_{\mathrm{D}}$ and $R_{\mathrm{Li}}$ as a function of 
$\sigma $. In standard (homogeneous) BBN, $^{4}$He varies much more slowly
with $\eta $ than do D and $^{7}$Li \cite{Cyburt}. Consequently, the change
in $^{4}$He is negligible for the range of $\sigma $ values considered here;
we find that $Y_{p}$ is altered by less than 0.1\%.

\begin{figure}[tbh]
\centerline{\epsfxsize=6truein\epsffile{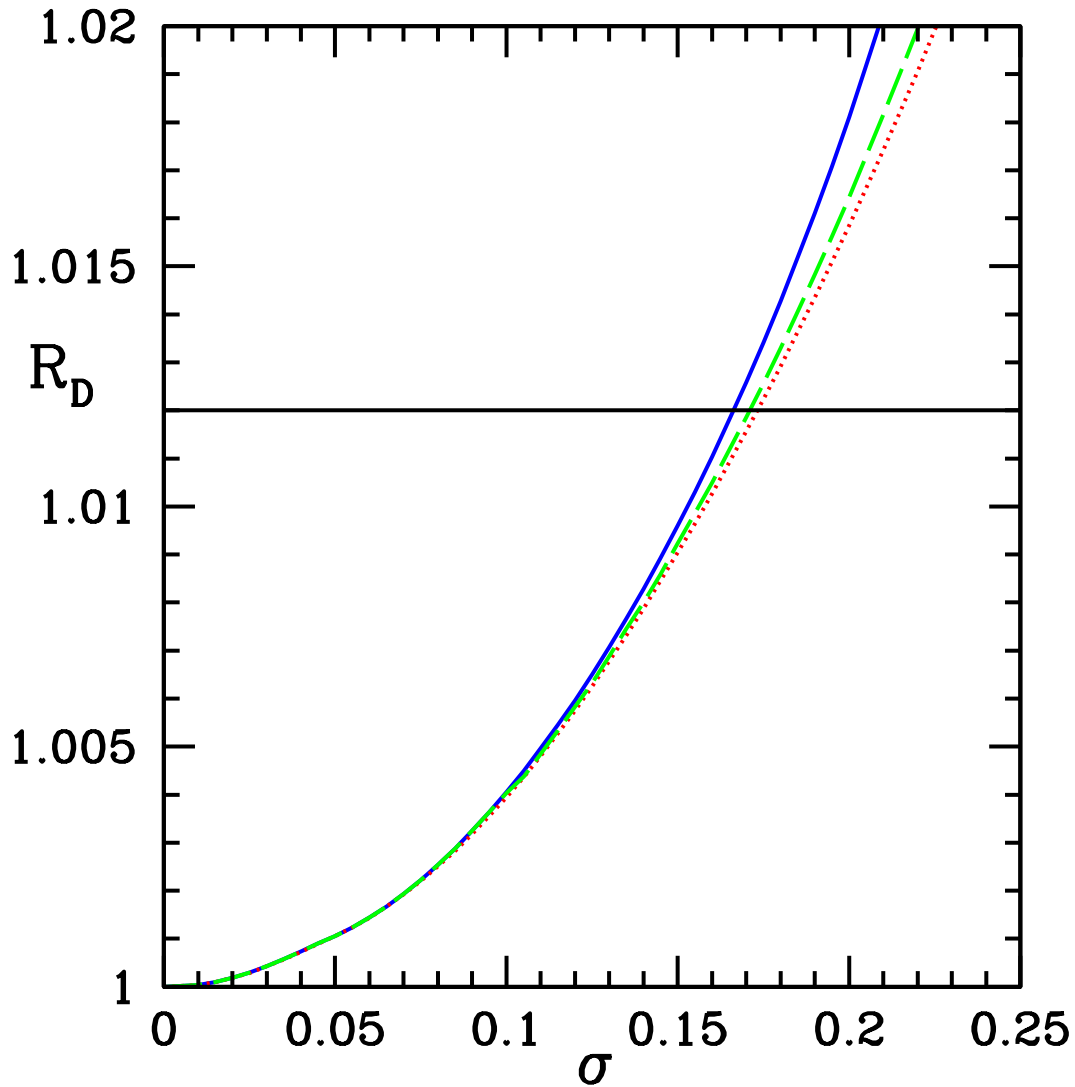}}
\caption{The ratio $R_{\mathrm{D}}$ of the value of D/H for inhomogeneous
BBN to the value of D/H in standard homogeneous BBN as a function of $
\protect\sigma $ (the RMS fluctuation in $\protect\eta /\overline{\protect
\eta }$) for the gaussian distribution (blue, solid), the lognormal
distribution (red, dotted), and the gamma distribution (green, dashed).
Horizontal line gives the $2-\sigma$
observational upper bound on this ratio derived from Ref. 
\protect\cite{PDG}.}
\end{figure}

\begin{figure}[tbh]
\centerline{\epsfxsize=6truein\epsffile{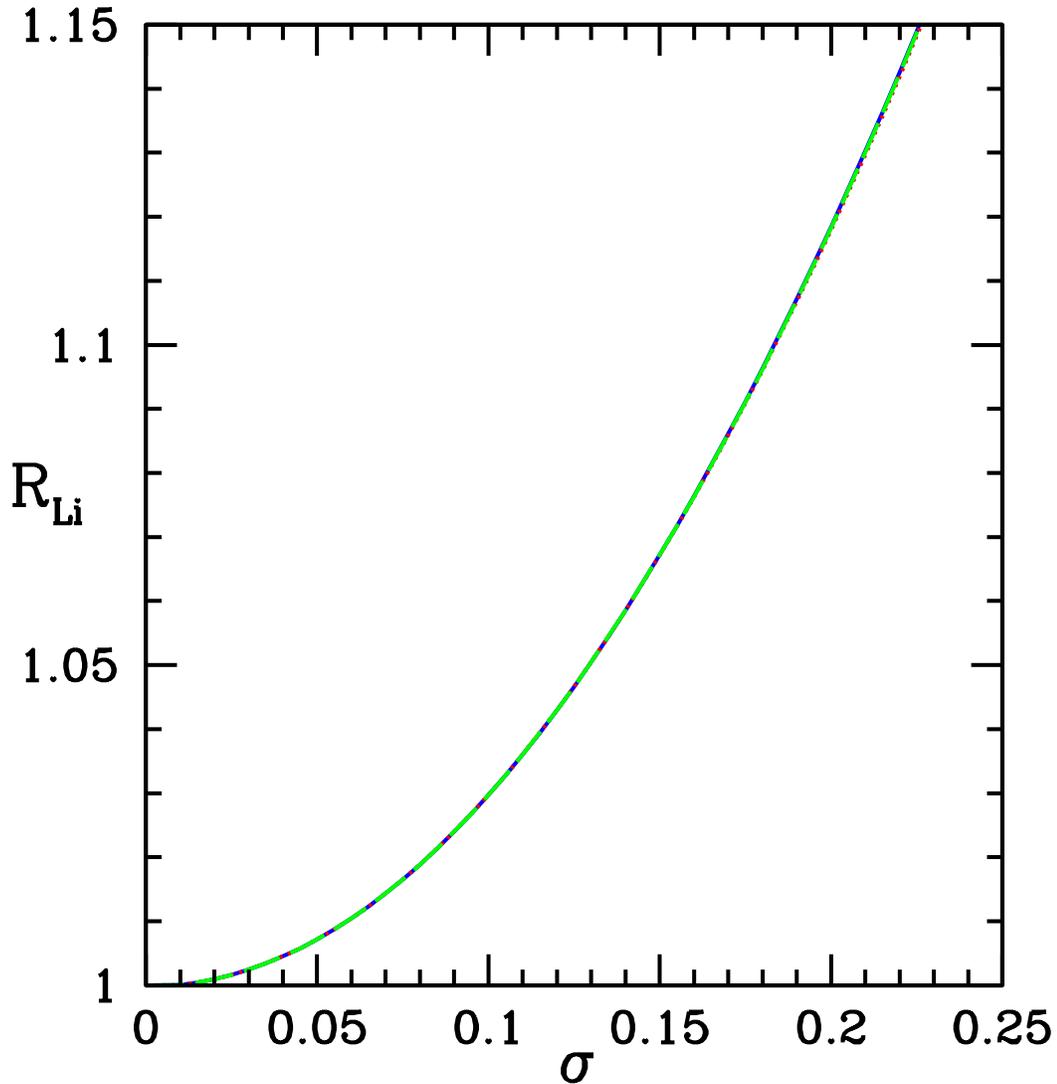}}
\caption{The ratio $R_{\mathrm{Li}}$ of the value of $^{7}$Li/H for
inhomogeneous BBN to the value of $^{7}$Li/H in standard homogeneous BBN as
a function of $\protect\sigma $ (the RMS fluctuation in $\protect\eta /
\overline{\protect\eta }$) for the gaussian distribution (blue, solid), the
lognormal distribution (red, dotted), and the gamma distribution (green,
dashed).}
\end{figure}

Our results for D and $^{7}$Li are displayed in Figs. 1-2. We first make
some general observations. The effect of the inhomogeneities is to increase
the abundance of deuterium and $^{7}$Li for all three of our choices for $
f(\phi )$. Further, the specific choice for the distribution function $
f(\phi )$ has only a small effect: the variation of the element abundances
with $\phi $ is very similar for all three of our choices for $f(\phi )$. In
the limit where $\sigma \rightarrow 0$, all three of our distributions give
nearly identical results; this is because both the gamma and lognormal
distributions approach a gaussian distribution in this limit.

It is possible to derive accurate analytic estimates for $R$ 
in the small-$\sigma$ limit.  Ref. \cite{Cyburt} provides approximations
for the element abundances as a function of $\eta$ when $\eta$ is near the standard model
value ($\eta = 6.1 \times 10^{-10}$).  Reexpressing these abundances in terms of $\phi$, we have \cite{Cyburt}
\begin{eqnarray}
\label{Dfit}
{\rm D/H} &=& ({\rm D/H})_{\phi=1} \phi^{-1.60},\\
\label{Lfit}
^7{\rm Li/H} &=& (^7{\rm Li/H})_{\phi=1} \phi^{2.11},
\end{eqnarray}
where the $\phi=1$ subscript denotes the predicted element abundances at $\phi=1$ ($\eta = 6.1 \times
10^{-10}$).  Assuming that $\overline Y_p$
varies very little from its homogeneous value (a good approximation),
we can substitute $X(\phi) = X_{\phi=1} \phi^n$ along with
the gaussian expression for $f(\phi)$ (Eq. \ref{gaussian}) into Eq. (\ref{Aphi}) and expand
to second order in $\sigma$ to obtain
\begin{equation}
\label{approx}
R \approx 1 + \frac{1}{2}n(n+1)\sigma^2,
\end{equation}
where $n = -1.60$ for deuterium and $n=2.11$ for $^7$Li.  Eq. (\ref{approx}) provides an excellent
approximation to $R_{\rm D}$ and $R_{\rm Li}$ for small values of $\sigma$.  While this analytic argument
assumes a gaussian distribution, our other two distributions, as we have noted, approach a gaussian
in the limit of small $\sigma$, so Eq. (\ref{approx}) applies to them as well in the small-$\sigma$ limit.

Eq. (\ref{approx}) shows that $R-1$ increases quadratically with $\sigma$ in the small-$\sigma$ limit, but
we can derive a more general result that is valid even when Eqs. (\ref{Dfit}) and (\ref{Lfit}) are
no longer good approximations. The quantity multiplying $f(\phi)$ in Eq. (\ref{Aphi}) is $X_A(\phi)\phi$.
When Eqs. (\ref{Dfit})-(\ref{Lfit}) are valid, this quantity is,
respectively, $\phi^{-0.60}$ for deuterium, and $\phi^{3.11}$ for
$^7$Li.  These are both convex functions (i.e., with positive second derivative). 
However, $X_A(\phi) \phi$ for both $^7$Li and deuterium
remains a convex function beyond the range of validity of Eqs.
(\ref{Dfit})-(\ref{Lfit}).  Whenever this quantity is convex, Jensen's inequality applied
to Eq. (\ref{Aphi}) implies that $R > 1$, i.e., the effect of inhomogeneities
is to increase the
deuterium and $^7$Li abundances relative to their homogeneous abundances.
Furthermore, this result is
independent of the functional form for $f(\phi)$ as long as $f(\phi)$ is small outside
the range where $X_A(\phi) \phi$ is convex.
As a corollary, the
kinds of inhomogeneities we are considering here cannot solve the lithium problem, since a solution
of that problem requires that
the $^7$Li abundance predicted by BBN be decreased, not increased.

Not surprisingly, the observed deuterium abundance gives the best upper bound on $\sigma $. Using
the $2-\sigma$ upper limit on D/H from the observational estimate in
Eq. (\ref{D/H}) and taking D/H = $
2.592\times 10^{-5}$ for the theoretical value in the homogeneous case at $
\phi =1$, we obtain an upper bound on $R_{D}$: 
\begin{equation}
R_{\mathrm{D}} < 1.012.
\end{equation}
This limit is displayed in Fig. 1. While there is some small variation
between the results for our three distributions, a conservative
bound derived from the observational limit is 
\begin{equation}
\sigma  < 0.17,
\end{equation}
i.e., the RMS fluctuation in $\eta $ must be smaller than $17\%$ of the mean value of $\eta$.

Lithium-7 increases more sharply with $\sigma $ in the inhomogeneous case
than does deuterium. However, as we have noted, BBN already predicts a
primordial $^{7}$Li abundance much larger than the observationally-estimated
primordial abundance, so our $^{7}$Li results cannot be used to place useful
limits on inhomogeneous BBN. The best we can do is to note that this model
cannot ameliorate the primordial lithium problem; indeed, it makes the
problem worse. 

\section{Discussion}

It is clear that the combination of sharp upper bounds on the deuterium
abundance, along with CMB limits on $\eta $, allows us to place tight
constraints on the graininess of cosmological models of the early universe
with large-scale baryon inhomogeneities. For all of the models examined
here, the RMS fluctuation in $\eta $ is constrained to be less than $17\%$
of the mean value of $\eta $. These constraints are due entirely to the
upper bound on deuterium. The $^{4}$He abundance is much less sensitive to
inhomogeneities in $\eta $, while the observationally-inferred primordial $
^{7}$Li abundance is already in conflict with the predictions of standard
BBN and so cannot constrain variations to BBN. Note further that the effect
of inhomogeneities is to increase the $^{7}$Li abundance, so inhomogeneous
BBN cannot provide a solution to the primordial lithium problem.

The obvious generalization of this work would be a reconsideration of
fluctuations on smaller scales, where differential neutron and proton
diffusion becomes important \cite{Applegate,Alcock,Lara,Nakamura}. Such models,
however, are considerably more complex, since the geometry and magnitude of
the fluctuations both have a strong influence on the final results.

\section{Acknowledgments}

JDB is supported by the Science and Technology Facilities Council (STFC) of the UK.

\end{document}